\documentclass{emulateapj}

\begin{document}

\shortauthors{Luhman et al.}
\shorttitle{Edge-on Disk around a Brown Dwarf}

\title{{\it Hubble} and {\it Spitzer} Observations of an Edge-on Circumstellar
Disk around a Brown Dwarf\altaffilmark{1}}

\author{
K. L. Luhman\altaffilmark{2,3},
Luc\'\i a Adame\altaffilmark{4},
Paola D'Alessio\altaffilmark{5}, 
Nuria Calvet\altaffilmark{6}, 
Kim K. McLeod\altaffilmark{7},
C. J. Bohac\altaffilmark{8},
William J. Forrest\altaffilmark{8},
Lee Hartmann\altaffilmark{6}, 
B. Sargent\altaffilmark{8},
\& Dan M. Watson\altaffilmark{8}
}

\altaffiltext{1}{
Based on observations made with the NASA/ESA {\it Hubble Space
Telescope} and the {\it Spitzer Space Telescope}.
The {\it Hubble} observations are associated with proposal ID 10511 and 
were obtained at the Space Telescope Science Institute, which is operated by 
the Association of Universities for Research in Astronomy, Inc., under NASA 
contract NAS 5-26555. {\it Spitzer} is operated by the Jet Propulsion 
Laboratory at the California Institute of Technology under NASA contract 1407.}

\altaffiltext{2}{Department of Astronomy and Astrophysics,
The Pennsylvania State University, University Park, PA;
kluhman@astro.psu.edu.}

\altaffiltext{3}{Visiting Astronomer at the Infrared Telescope Facility, which
is operated by the University of Hawaii under Cooperative Agreement no. NCC
5-538 with the National Aeronautics and Space Administration, Office of Space
Science, Planetary Astronomy Program.}

\altaffiltext{4}{Instituto de Astronom\'\i a, Universidad Nacional
 Aut\'onoma de M\'exico, Apartado Postal 70-264, 
Ciudad Universitaria, M\'exico DF, M\'exico.}

\altaffiltext{5}{Centro de Radioastronom\'\i a y Astrof\'\i sica, 
Universidad Nacional Aut\'onoma de M\'exico, Apartado Postal 72-3 (Xangari), 
Morelia, Michoac\'an, M\'exico.}

\altaffiltext{6}{Department of Astronomy, The University of Michigan, 
500 Church Street, 830 Dennison Building, Ann Arbor, MI.}

\altaffiltext{7}{Whitin Observatory, Wellesley College, Wellesley, MA.}

\altaffiltext{8}{Department of Physics and Astronomy, 
The University of Rochester, Rochester, NY.}

\begin{abstract}

We present observations of a circumstellar disk 
that is inclined close to edge-on around a young brown dwarf 
in the Taurus star-forming region. 
Using data obtained with SpeX at the NASA Infrared Telescope Facility, 
we find that the slope of the 0.8-2.5~\micron\ spectrum of the brown dwarf
2MASS~J04381486+2611399 cannot be reproduced with a photosphere reddened by 
normal extinction. Instead, the slope is consistent with scattered light, 
indicating that circumstellar material is occulting the brown dwarf. 
By combining the SpeX data with mid-infrared photometry and 
spectroscopy from the {\it Spitzer Space Telescope} and previously published
millimeter data from Scholz and coworkers, we construct the spectral energy
distribution for 2MASS~J04381486+2611399 and model it in terms of a young 
brown dwarf surrounded by an irradiated accretion disk. 
The presence of both silicate absorption at 10~\micron\ and silicate emission
at 11~\micron\ constrains the inclination of the disk to be $\sim70\arcdeg$,
i.e. $\sim20\arcdeg$ from edge-on.
Additional evidence of the high inclination of this disk is provided by
our detection of asymmetric bipolar extended emission surrounding 
2MASS~J04381486+2611399 in high-resolution optical images obtained with 
the {\it Hubble Space Telescope}.
According to our modeling for the SED and images of this system, the disk
contains a large inner hole that is indicative of a transition disk
($R_{in}\approx58R_\star\approx0.275$~AU) and is somewhat larger than expected 
from embryo ejection models ($R_{out}=20$-40~AU vs. $R_{out}<10$-20~AU).

\end{abstract}

\keywords{accretion disks -- planetary systems: protoplanetary disks -- stars:
formation --- stars: low-mass, brown dwarfs --- stars: pre-main sequence}

\section{Introduction}
\label{sec:intro}

Measurements of the properties of circumstellar accretion disks around young 
stars are important because they represent constraints on the initial 
conditions of planet formation. 
A unique set of measurements can be performed on a disk in the rare case
in which it is seen close to edge-on.
Because an edge-on disk occults the central star, scattered light
from the disk surface dominates the total emergent flux at optical and 
near-infrared (IR) wavelengths, making it possible to spatially resolve the disk
with high-resolution imaging \citep{bur96}. At mid-IR wavelengths, the outer
disk is seen in absorption against the star and inner disk, allowing 
spectroscopic measurements of the disk composition \citep{wat04,pon05}.
The sample of edge-on disks discovered to date remains small. 
Some of the most notable examples are
Orion~114-426 \citep{mcc96}, Haro~6-5B \citep{kri98}, HH~30 \citep{bur96}, 
IRAS~04302+2247 \citep{luc97,pad99}, DG~Tau~B \citep{pad99},
OphE-MM3, CRBR 2422.8-3423 \citep{bra00}, HK~Tau~B \citep{sta98,kor98}, 
HV~Tau~C \citep{mon00}, LkH$\alpha$~263~C \citep{jay02,cha02},
2MASS~J1628137-243139 \citep{gro03}, and PDS~144 \citep{per06}.
Additional young stars in Taurus exhibit dust lanes that may also
trace edge-on disks \citep{har99,pad99}.

Because young stars with edge-on disks are seen only in scattered light
at optical and near-IR wavelengths,
they appear much fainter than unocculted young stars at a given spectral type. 
Recent spectroscopic surveys of nearby star-forming regions have identified
a number of objects that are subluminous in this manner 
\citep{fer01,luh03,sle04}.  
One of these sources, 2MASS~J04381486+2611399 in Taurus (henceforth
2M~0438+2611), appears to be a brown dwarf with a mass near 0.05~$M_\odot$ 
\citep{luh04tau}.
This brown dwarf also exhibits anomalous near-IR colors and strong forbidden 
line emission, which are characteristics that are frequently observed in stars
with edge-on disks.
To definitively establish whether 2M~0438+2611 has an edge-on disk, 
we have observed it with near- and mid-IR spectroscopy and high-resolution 
optical imaging.
In this paper, we present these new data (\S~\ref{sec:obs}) 
and fit them with the predictions of accretion disk models to constrain 
the inclination and other physical properties of the disk around 2M~0438+2611
(\S~\ref{sec:model}).

\section{Observations and Analysis}
\label{sec:obs}

\subsection{Near-infrared Spectroscopy}
\label{sec:spex}

To investigate the possibility that 2M~0438+2611 is seen in scattered light,
we first obtained a low-resolution near-IR spectrum of it with SpeX 
\citep{ray03} at the NASA Infrared Telescope Facility (IRTF). These data
were collected on the night of 2004 November 12. They were reduced with the 
Spextool package \citep{cus04} and corrected for telluric absorption 
\citep{vac03}. The final spectrum extends from 0.8-2.5~\micron\ and 
exhibits a resolving power of $R=100$.

\begin{figure}
\plotone{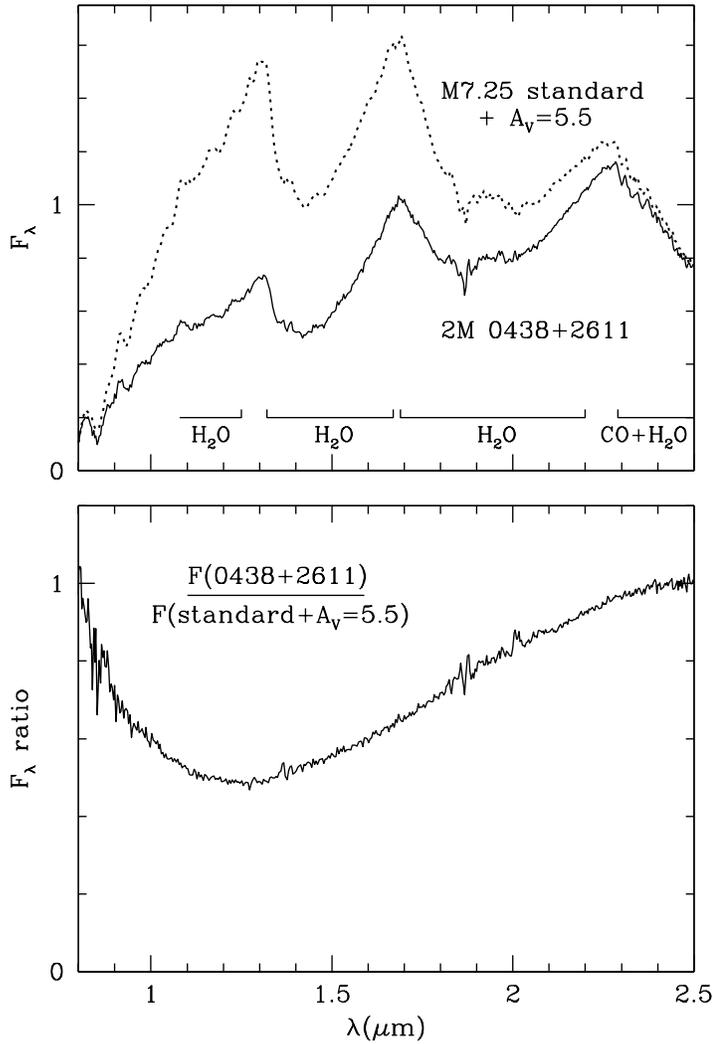}
\caption{
{\it Top}:
Near-infrared spectrum of the young brown dwarf 2M~0438+2611 ({\it solid line}) 
compared to data for a standard young brown dwarf with the same optical 
spectral type ({\it dotted line}).
The spectrum of 2M~0438+2611 is normalized at 1.68~\micron.  The standard 
spectrum has been reddened by $A_V=5.5$ so that it has the same relative
fluxes at 0.8 and 2.5~$\micron$ as 2M~0438+2611. The overall shape of
2M~0438+2611 cannot be reproduced by applying reddening to the standard.
{\it Bottom}: The anomalous slope of 2M~0438+2611 is further
illustrated by the ratio of the spectra of these two objects
({\it solid line}), which departs significantly from the constant ratio
expected for normal reddening of a photosphere. The absence of 
significant residuals in this ratio near the steam bands confirms
that the two objects have similar spectral types.}
\label{fig:spec}
\end{figure}

For most of the young brown dwarfs that we have observed in previous studies
\citep[e.g.,][]{luh06tau2},
the differences in the slopes of their 0.8-2.5~\micron\ spectra are 
consistent with differences in extinction. However, this is not the case for
2M~0438+2611, which has an anomalous slope that differs
significantly from those of other objects, regardless of any correction for
extinction using standard reddening laws \citep{rl85,car89}.
This behavior is illustrated in Figure~\ref{fig:spec},
where we compare 2M~0438+2611 to a typical
unreddened young brown dwarf with the same optical spectral 
type\footnote{For the standard spectrum, we use SpeX data for 
2MASS~04442713+2512164 \citep[henceforth 2M~0444+2512,][]{luh04tau} after 
dereddening it by $A_V=2.3$, which is the extinction implied by a comparison to 
other young brown dwarfs and to dwarf standards. Based on that comparison,
the SpeX data for 2M~0444+2512 do not exhibit excess emission from a disk.}
The standard was artificially reddened according to 
the reddening law of \citet{car89} to the point that it has the same relative
fluxes at 0.8 and 2.5~$\micron$ as 2M~0438+2611. However, the shape of this
reddened spectrum between 0.8 and 2.5~$\micron$ differs significantly from 
that of 2M~0438+2611.
An alternative demonstration of this effect is shown in 
Figure~\ref{fig:spec} through the quotient these two spectra, which departs
from the constant value expected for normal reddening.
Thus, reddening cannot explain the observed slope of 2M~0438+2611.
Instead, relative to the flux at 0.8~\micron, the spectrum of 2M~0438+2611 
becomes redder more slowly with longer wavelengths than expected from standard 
extinction laws, which is 
consistent with the presence of (blue) scattered light in the spectrum.
Meanwhile, the absence of significant residuals in the quotient of 2M~0438+2611 
and the standard demonstrates the close match in the depths of the steam 
absorption bands, supporting the similarity in spectral types indicated by 
previous optical spectroscopy. 

\subsection{Mid-infrared Spectroscopy}

Photometry at 2~\micron\ from the Two-Micron All-Sky Survey 
\citep[2MASS,][]{skr06}, at 3-8~\micron\ from the Infrared Array
Camera \citep[IRAC;][]{faz04} aboard the {\it Spitzer Space Telescope}
\citep{wer04}, and at 1.3~mm from the 30~m telescope at the Institut de Radio 
Astronomie Millim{\'e}trique (IRAM)
have previously revealed excess emission in 2M~0438+2611
that is indicative of circumstellar dust \citep{luh04tau,luh06tau2,sch06}.
To better constrain the properties of this material,
we obtained a mid-IR spectrum of 2M~0438+2611 on 2005 March 19
with the {\it Spitzer} Infrared Spectrograph \citep[IRS;][]{hou04} as a part 
of the Guaranteed Time Observations of the IRS instrument team.  
These observations (AOR 12705792) were performed with the short-wavelength, 
low-resolution module of IRS, providing data from 5.3-14~\micron\ with a 
resolving power of $R=90$. The spectrum was processed with the S14 
pipeline at the {\it Spitzer} Science Center. The remaining reduction 
was performed with the methods that have been previously 
applied to IRS data for other low-mass members of Taurus \citep{fur05}.
We have also measured photometry at 24~\micron\ for 2M~0438+2611 from
archival images obtained with the Multiband Imaging Photometer for
{\it Spitzer} \citep[MIPS;][]{rie04}. Using the methods described
by \citet{all07}, we measured a flux of $62.5\pm3.3$~mJy from these MIPS data.

We present the spectra from SpeX and IRS and the photometry from 2MASS, IRAC,
MIPS, and IRAM for 2M~0438+2611 in Figure~\ref{fig:sed}. 
The IRAC and IRS data agree well with each other, 
while the 2MASS photometry and the SpeX data differ significantly in 
both color and flux level.  The SpeX data are consistent with a
smooth extension of the IRAC and IRS data, while the 2MASS measurements
appear to be discontinuous from the latter. 
The IRAC and IRS observations were performed only 4 months after the 
SpeX observations, while the 2MASS data were obtained 6 years earlier.
Thus, one explanation for the discrepancy in the 2MASS photometry relative
to the other data is variability, which is plausible for any young star
and is particularly likely for an object that is seen in scattered light
because of changes in the geometry of the occulting material.
Therefore, we exclude the 2MASS photometry for the purposes of modeling 
the spectral energy distribution (SED) of 2M~0438+2611 in \S~\ref{sec:model}.

\subsection{Optical Imaging}
\label{sec:wfpc2}

Because of the initial evidence indicating that the SED of 2M~0438+2611 might
be dominated by scattered light at optical and near-IR wavelengths 
\citep[][\S~\ref{sec:spex}]{luh04tau}, we sought to detect spatially resolved 
scattered light through high-resolution broad-band optical imaging with the 
{\it Hubble Space Telescope}. In addition, given the presence of 
forbidden line emission in optical spectra of this brown dwarf \citep{luh04tau},
we performed narrowband imaging centered on [O~I] at 6300~\AA\ in an attempt 
to detect resolved line emission from a jet or an outflow. Using the Wide Field 
Planetary Camera (WFPC2) aboard {\it Hubble}, we obtained images of
2M~0438+2611 through the F675W, F791W, F850LP, and F631N filters
on 2005 October 22. The target was placed near the center of the PC, which
has a plate scale of $0.046\arcsec$~pixel$^{-1}$. Two images were taken in each 
filter, each with exposure times of 160~sec for F675W, F791W, and F850LP and 
300~sec for F631N. Each pair of images at a given filter was combined 
to create a single image.

In the [O~I] image, 2M~0438+2611 is unresolved and no extended emission 
is detected. Meanwhile, each of the three broad-band images reveal 
both a point source and spatially resolved emission, as shown in 
Figure~\ref{fig:image}. The extended emission is elongated and reaches 
0.4-$0.5\arcsec$ from one side of the point source at a position angle of 
$245\arcdeg$.  Analysis of these images with the point-spread-function of WFPC2 
indicates that a small amount ($\sim0.1\arcsec$) of extended emission is present
on the opposite side of the point source as well.

\section{Disk Model}
\label{sec:model}

\subsection{Model Parameters}
\label{sec:para}

We have modeled the SED of 2M~0438+2611 in Figure~\ref{fig:sed} following the 
procedures from \citet{dal98,dal99,dal01,dal06}. 
In short, we solve the equations for the disk vertical structure, assuming 
it is an $\alpha$-disk heated by viscous dissipation and by stellar 
irradiation. The relatively high flux at 1.3~mm of this object 
\citep{sch06} suggests the presence of grains that are larger than typical 
grains in the interstellar medium. At the same time, the presence of the 
10~\micron\ silicate band and the shape of the extinction of the stellar SED
are indicative of small grains. Therefore, we have constructed a disk 
model in which a large fraction of the dust has settled in the midplane,
growing to a maximum size of $\sim1$~mm, while a small fraction of the
dust remains in the upper layers in the form of small interstellar-like 
grains \citep{dal06}. 
The dust is assumed to be segregated spheres of ``astronomical'' silicates
and graphite with abundances and optical constants from \citet{dl84}
and \citet{wl01} and a size distribution of $n(a)\sim a^{-3.5}$ where 
$a$ is the grain radius \citep{mrn}.
For the small grains in the upper layers, the minimum and maximum grain
radii are $a_{min}=0.0005$~\micron\ and $a_{max}=0.25$~\micron.
For the large grains at the disk midplane, we use 
$a_{min}=0.0005$~\micron\ and $a_{max}=1$~mm. The dust-to-gas mass
ratio of the small grains is parameterized in terms of $\epsilon$,
which is the ratio normalized by the standard interstellar value of $\sim0.01$. 
We have calculated models for $0.01<\epsilon<1$. The dust-to-gas mass ratio 
of the large grains at the midplane is calculated assuming that the total mass 
in grains is conserved at each radius.

In our model, the dusty disk is truncated at a radius $R_{in}$, where the
inner wall of the disk receives radiation from the brown dwarf and accretion
shocks at the stellar surface with a normal incidence. A natural
explanation for a wall of this kind is that the silicates are sublimated
inside $R_{in}$ \citep{dul01,muz04}. In this case, the temperature at the 
inner edge of the disk is $T_{wall}=T(R_{in})\sim1400$~K.  
However, disks can be truncated at larger radii, corresponding to lower 
temperatures (i.e., transitional disks). Therefore, we have modeled the wall
emission following the procedures of \citep{dal05}, varying the temperature 
of the optically thin dust in the wall from 300 to 1400~K.
We have adopted an accretion rate of $3\times10^{-11}$~$\rm M_{\odot}\,yr^{-1}$
for our disk model, which is similar to the value derived by \citet{muz05} for
2M~0438+2611 through modeling of the profile of its H$\alpha$ emission line.

Although the disk surface density, $\Sigma$, is not an input parameter for our 
models, we are able to modify it through the viscosity parameter $\alpha$ since 
$\Sigma\sim\frac{\dot{M}}{\alpha}$ for an $\alpha$-disk. 
When fitting the millimeter flux, decreasing $\alpha$ (i.e., increasing
$\Sigma$) has the effect of decreasing the outer radius of the disk.
After exploring models for $10^{-6}\leq\alpha\leq10^{-2}$,
we find that the observed millimeter flux constrains $\alpha$ to the low 
end of this range if the accretion rate in the outer disk is the same as 
the accretion rate onto the brown dwarf \footnote{Disks around T Tauri stars 
have been modeled with $\alpha\sim0.01$ \citep{dal98}.}.
For a brown dwarf disk, $\alpha\leq10^{-4}$ implies a viscous 
timescale of $t_\nu\gtrsim25$~Myr, which is too long to justify our
assumption of a steady disk with a constant accretion rate throughout the disk.
It is possible that the accretion rate increases with disk radius,
as in the disk model from \citet{gam96}. A higher accretion rate in the
outer disk would correspond to lower $\alpha$ and shorter $t_\nu$. 
In a disk of this kind, material would accumulate in the inner disk, 
perhaps in a dead zone. Exploring this possibility in detail would require
a disk model that includes a dead zone.

For the photosphere of 2M~0438+2611, we have adopted an effective temperature 
of 2838~K \citep{luh04tau}. Because this object is seen in scattered light 
at optical and near-IR wavelengths \citep[][\S~\ref{sec:spex}]{luh04tau}, 
its extinction cannot be measured from its colors. As a result, 
we cannot reliably measure its luminosity with the normal method of 
applying an extinction correction to photometry in these bands. 
Therefore, we have performed the disk calculations for a range
of luminosities that are typical of members of Taurus near the spectral
type of 2M~0438+2611, namely $L_{\rm bol}=0.04$, 0.06, 0.08, and 0.1~$L_\odot$. 
For 2M~0438+2611, we adopt a mass of 0.05~$M_\odot$, which is the value 
implied by its spectral type for a member of Taurus \citep{luh04tau}.
As shown in \citet{luh04tau} and \S~\ref{sec:spex}, 2M~0444+2512
has the same optical and IR spectral types as 2M~0438+2611. In addition, 
although it exhibits mid-IR excess emission that indicates the presence
of a disk \citep{luh06tau2}, 2M~0444+2512 does not show excess emission
at wavelengths shortward of 2.5~\micron\ in a comparison to SpeX data 
for diskless brown dwarfs near the same spectral type \citep{luh06tau1}. 
Therefore, we have adopted the extinction-corrected SpeX data for 2M~0444+2512
described in \S~\ref{sec:spex} to represent the 0.8-2.5~\micron\ SED of the 
photosphere of 2M~0438+2611.
We assume that both 2M~0438+2611 and 2M~0444+2512 are at the average 
distance of members of Taurus \citep[$d=140$~pc,][]{wic98}. 
We measured the average colors between 2MASS $K_s$ and the IRAC bands
(3.6, 4.5, 5.8, 8.0~\micron) for diskless late-type members of Taurus 
\citep{har05,luh06tau2} and applied them to the SpeX data for 2M~0444+2512
to extend the photospheric SED to 8.0~\micron. The SED was extrapolated to 
wavelengths beyond 8.0~\micron\ with a Rayleigh-Jeans distribution.

\subsection{Best-fit Model}

A given portion of the observed SED provides constraints on specific 
properties of the disk. 
We use the $K$-band flux to estimate the optical depth to the brown dwarf, 
which in turn constrains the inclination, outer radius, and viscosity parameter
of the disk. The silicate feature near 10~\micron\ is sensitive to both 
the inclination and the presence of small grains in the upper layers
of the disk. The abundance of these small grains is constrained
by the 24~\micron\ flux relative to the emission at shorter wavelengths. 
The flux at 3-10~\micron\ depends on the inner radius 
of the disk and the stellar luminosity, while the millimeter flux is 
determined by a combination of the outer radius and the surface density.
In this section, we discuss each of these constraints in detail 
for 2M~0438+2611 and present the resulting best-fit model for its disk.

The best fit to the flux at 3-10~\micron\ is provided by a stellar luminosity
of 0.06~$L_\odot$ (corresponding to $R_*=1.02$~$R_\odot$) and a 
radius of $R_{in}\approx58R_\star\approx0.275$~AU for the wall, which 
corresponds to a wall temperature of 400~K.
For models with the wall placed at the dust destruction radius, the predicted 
fluxes at 3-10~\micron\ are brighter than the observed values. 
Thus, our modeling indicates that the disk is inwardly 
truncated. The inner disk of 2M~0438+2611 shows the same physical
properties as disks that have been identified in the
literature as ``transitional disks'', and which have been
shown to have truncated optically thick disks at various
distances from the central stars, from both SED modeling
\citep{cal02,cal05,dal05,muz06} and millimeter interferometry \citep{hug07}.
Disks of this kind are thought to harbor forming planets that are 
opening gaps in the disk \citep{ric03,qui04}. Photoevaporation is an
alternative explanation for these disk gaps, but it is unlikely for
2M~0438+2611 given the low ultraviolet fluxes that are expected from a
brown dwarf \citep{muz06}. 

For all reasonable choices of model parameters, the reddened flux from the 
stellar photosphere should dominate the total emergent flux at 
$\sim2$-3~\micron. Combining the luminosity of 0.06~$L_\odot$ and the 
observed $K$-band flux produces a value of $\tau_{2.25\mu m}\approx1.3$ 
for the optical depth to the photosphere.

\begin{figure}
\epsscale{.8}
\plotone{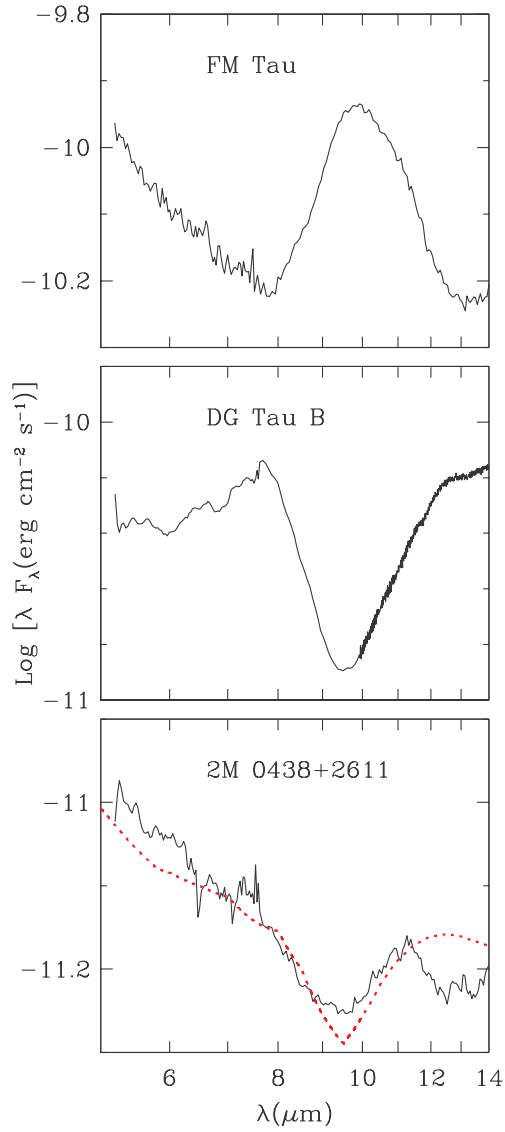}
\caption{
Mid-infrared spectrum of the young brown dwarf 2M~0438+2611 compared
to data for a face-on disk \citep[FM~Tau,][]{fur06} and 
an edge-on disk \citep[DG~Tau~B,][]{wat04}, which exhibit silicate 
emission and absorption, respectively. 
To reproduce both the emission and the absorption observed in the spectrum of 
2M~0438+2611, our models for its disk ({\it dotted line}) require an 
inclination near $\sim70\arcdeg$ (i.e., $20\arcdeg$ from edge-on).
The disk model shown here is for $R_{out}=40$.
}
\label{fig:sed2}
\end{figure}

The 10~\micron\ silicate profile of 2M~0438+2611 is distinctive, showing
both absorption and emission components. This profile tightly constrains
the disk inclination because small angles produce only emission and large
ones produce only absorption, as illustrated for the face-on and edge-on 
disks around FM~Tau and DG~Tau~B in Figure~\ref{fig:sed2} \citep{wat04,fur06}.
For $i\sim70\arcdeg$, the model predicts silicate emission from the disk 
and silicate absorption from the highly extinguished wall that combine
to form absorption and emission features at 10 and 11~\micron, which closely 
matches the data, as shown in Figure \ref{fig:sed2}.
While modeling IR and millimeter photometry for 2M~0438+2611, \citet{sch06}
also used a high inclination angle of $i\sim80\arcdeg$ for their model of
this disk. A high inclination was produced by their model because
2M~0438+2611 is much fainter than the photosphere of typical young brown
dwarfs, so a highly inclined, obscuring disk was needed to suppress the
near- and mid-IR fluxes of their adopted photospheric template 
to the observed levels. Thus, their evidence for a high
inclination disk was equivalent to that presented by \citet{luh04tau}, 
who showed that 2M~0438+2611 is anomalously faint at near-IR wavelengths for 
its spectral type and might be occulted by circumstellar material.  
The distinctive silicate features in our IRS spectra 
(as well as the extended emission in the WFPC2 images) represent 
new evidence of a highly-inclined disk around 2M~0438+2611. 
\citet{sch06} suggested that the disk around 2M~0444+2512
also might have a high inclination. However, the slope of its 
near-IR spectrum is consistent with a small amount of extinction and
a normal reddening law, and thus does not indicate the presence of scattered
light. In addition, an unpublished IRS spectrum of 2M~0444+2512 does
not show the silicate absorption that is seen in 2M~0438+2611  
and other highly-inclined disks (Figure~\ref{fig:sed2}).

The flux at 24~\micron\ is sensitive to the degree of dust settling in the disk.
We find that small grain abundances of $0.03<\epsilon<0.09$ in the upper layers
of the disk are required to explain the observed 24~\micron\ flux.
For values of $\epsilon$ that are outside of this range, 
the predicted slope of the mid- and far-IR SED is smaller or larger than the 
observed one. 

\begin{figure}
%\epsscale{.7}
\plotone{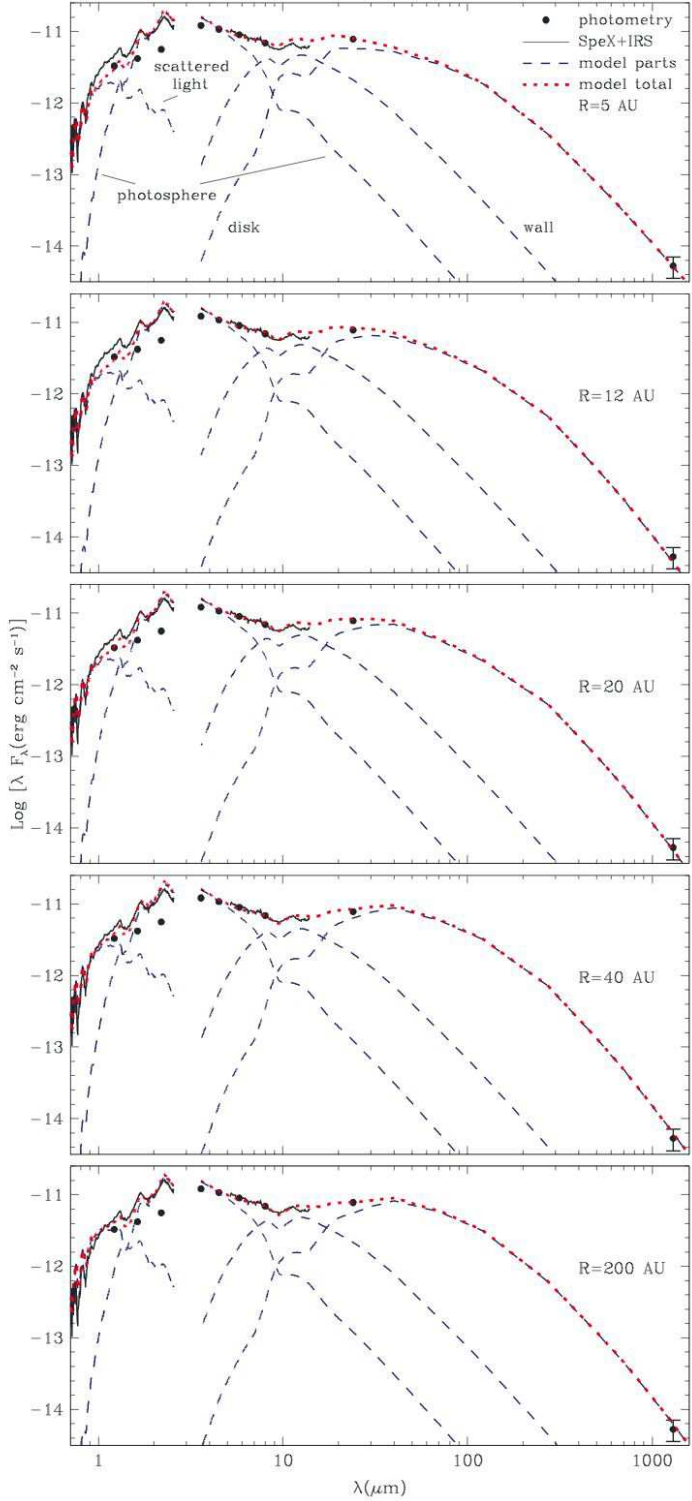}
\caption{
Spectral energy distribution of the young brown dwarf 2M~0438+2611 
compared to models for its circumstellar accretion disk. The data for 
2M~0438+2611 consist of photometry from 2MASS, IRAC, MIPS, and IRAM
({\it filled circles}) and spectroscopy from SpeX and IRS ({\it solid lines}).
The components of the model for 2M~0438+2611 consist of the stellar 
photosphere, light from the stellar photosphere scattered by the disk surface, 
emission from the inner disk wall, and emission from the disk 
({\it dashed lines}). 
We show the best fit models to the SED for $R_{out}=5$-200~AU
({\it top to bottom}, Table~\ref{tab:models}).
For each of these models, the sum of the model 
components ({\it dotted line}) agrees with most of the data, except for
the 2MASS photometry, which is discrepant from all of the other measurements.
The IRAC and IRAM data are from \citet{luh06tau2} and \citet{sch06},
respectively. All other measurements are from this work.
With the exception of the millimeter measurement, the uncertainties for 
the photometric measurements are smaller than the points.
}
\label{fig:sed}
\end{figure}

\begin{figure}
\epsscale{0.6}
\plotone{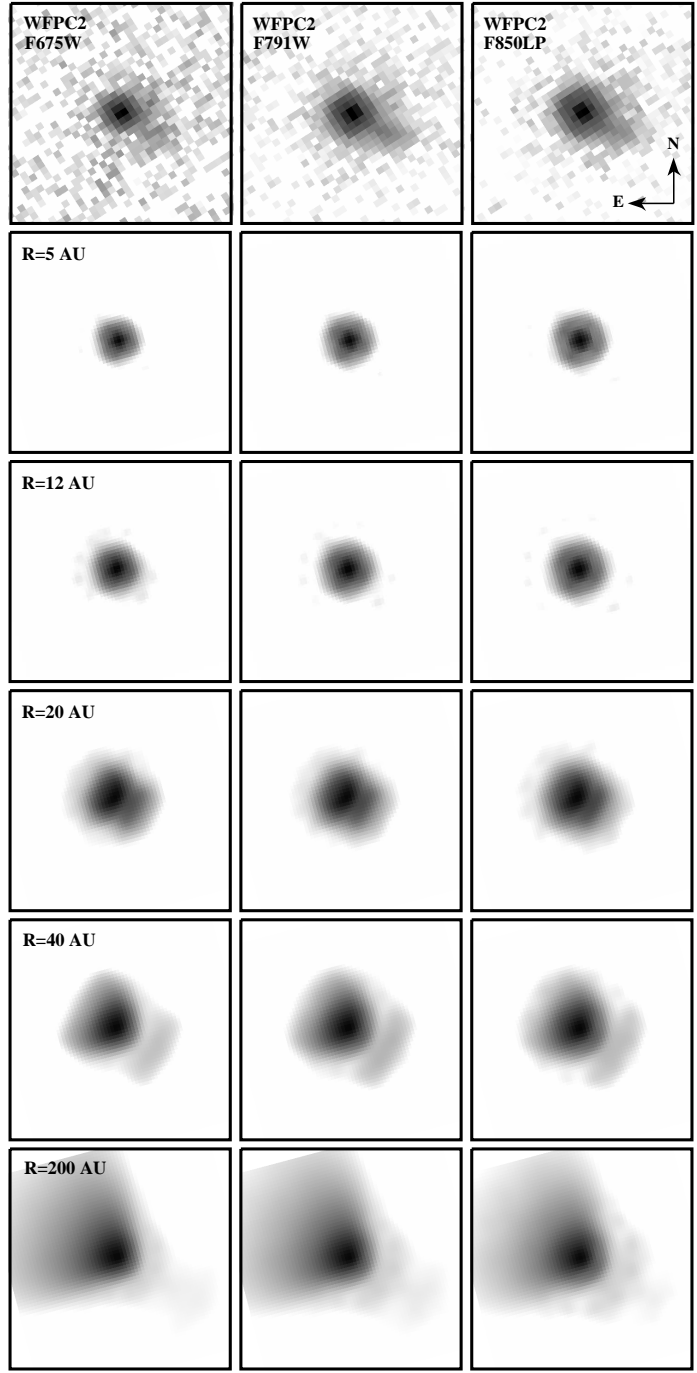}
\caption{
WFPC2 images of 2M~0438+2611 ({\it top row}) and simulated images produced by
the disk models that fit the SED in Figure~\ref{fig:sed} ({\it bottom
five rows}).
In the model images, the disk is close to edge-on ($i\sim70\arcdeg$) and is 
aligned close to north-south. Thus, the extended emission in the horizontal
direction is above and below the disk.
For each image, the intensity is displayed on a logarithmic scale
and the size is $1.5\arcsec\times1.5\arcsec$. 
}
\label{fig:image}
\end{figure}

In Table~\ref{tab:models}, we summarize the models that provide the best fits
to the SED of 2M~0438+2611 for five values of $\alpha$.
The SEDs produced by these models are shown with the observed SED in 
Figure~\ref{fig:sed}.
These models reproduce the observed SED longward of 2~\micron\ equally
well, while the model for $\alpha=10^{-4}$ provides a somewhat better match
to the flux at shorter wavelengths. The disk radii of these models range 
from 5 to 200~AU. Thus, the SED alone does not provide a
useful constraint on the outer radius of the disk. 

To estimate the disk radius for 2M~0438+2611, we compare the images produced by 
each of our five models to the WFPC2 images in Figure~\ref{fig:image}.
The WFPC2 data show a central peak surrounded by asymmetric bipolar emission.
In comparison, the model images for $R_{out}=5$ and 12~AU show a point 
source and no detectable extended emission and the model for 
$R_{out}=200$~AU produces too much extended emission.
Meanwhile, the images from the models for $R_{out}=20$ and 40~AU agree
reasonably well with the WFPC2 images of 2M~0438+2611.
The relative fluxes of the two lobes of emission are better matched by
the model for $R_{out}=20$~AU, while the length of the extended emission is 
better matched by $R_{out}=40$~AU. 
Thus, the WFPC2 images constrain the disk radius to be $R_{out}=20$-40~AU 
and confirm the high inclination that is implied by the SED analysis.
In addition, given that the model images are formed by stellar light scattered 
at the disk surface and extinguished by the outer disk, 
the agreement between the optical SED and the predicted scattered light 
spectrum and flux level (Figure~\ref{fig:sed}) is further evidence supporting 
our estimate of the depletion factor of small grains in the upper disk layers.

\section{Conclusions}

The young brown dwarf 2M~0438+2611 exhibits several characteristics that 
are often observed in stars with edge-on disks, such as unusually faint 
near-IR fluxes for its spectral type, strong emission in forbidden transitions, 
and anomalous near-IR colors \citep{luh04tau}. 
Through observations presented in this paper, 
we have confirmed that 2M~0438+2611 is occulted by a highly inclined disk.
This new evidence is summarized as follows:

\begin{enumerate}

\item
Based on a comparison to other young brown dwarfs,
the slope of the 0.8-2.5~\micron\ spectrum of 2M~0438+2611 
cannot be explained as a photosphere reddened by a standard extinction law
and instead is consistent with scattered light, which suggests that the 
brown dwarf is occulted by circumstellar material.

\item
The presence of silicate absorption at $\sim10$~\micron\ and 
silicate emission at $\sim11$~\micron\ constrains the disk inclination angle
to be near $\sim70\arcdeg$ (i.e., $20\arcdeg$ from edge-on) because higher or
lower inclinations would produce only absorption or emission, respectively.

\item
We detect asymmetric bipolar emission in WFPC2 images of 2M~0438+2611, which 
is consistent with simulated scattered-light images produced our model
of a highly-inclined disk around this brown dwarf. 

\end{enumerate}

In addition to its inclination, we have been able to constrain several other 
properties of the disk around 2M~0438+2611 through modeling of its SED
and high-resolution images.
These constraints are made possible by the unusual wealth of data available for 
this disk, although achieving a self-consistent model that simultaneously
reproduces those data has proved to be quite challenging.
In our best-fit models, the disk has an inclination of $\sim70\arcdeg$, 
an inner radius of $58R_\star\approx0.275$~AU, an outer
radius of 20-40~AU, a total mass of 100-200~$M_\oplus$, and an abundance of 
small grains in its upper layers of $0.03\lesssim\epsilon\lesssim0.09$, 
indicating a large degree of settling to the midplane.
Our estimate of the inner radius suggests the presence of a large inner hole,
which is a characteristic of transitional disks \citep{cal02,cal05,dal05,muz06}.
Meanwhile, the outer radius of the disk around 2M~0438+2611 is somewhat
larger than expected from models of embryo ejection 
\citep[$R_{out}<10$-20~AU,][]{bat03}.

\acknowledgements

We acknowledge support from 
grant GO-10511 from the Space Telescope Science 
Institute and grant AST-0544588 from the National Science Foundation (K. L.), 
grant 172854 from CONACyT (L. A.), grants from CONACyT and PAPIIT/DGAPA, 
M\'exico (P. D.), and NASA grants NAG5-9670 and NAG5-13210 (N. C., L. H.).

%\clearpage

\begin{deluxetable}{lllll}
\tablewidth{0pt}
\tablecaption{Disk Models for 2M~0438+2611\label{tab:models}}
\tablehead{
\colhead{$\alpha$} & 
\colhead{$R_{out}$} &
\colhead{$M_{disk}$} &
\colhead{$i$} &
\colhead{} \\
\colhead{($\times10^{-4}$)} &
\colhead{(AU)} &
\colhead{($M_\oplus$)} &
\colhead{(deg)} &
\colhead{$\epsilon$}}
\startdata
0.05 & 5 & 700 & 68 & 0.04 \\
0.5 & 12 & 140 & 70 & 0.03 \\
0.8 & 20 & 130 & 71 & 0.03 \\
1 & 40 & 190 & 67 & 0.09 \\
2 & 200 & 340 & 70 & 0.09 \\
\enddata
\end{deluxetable}

%\clearpage

%\clearpage

%\clearpage

%\clearpage

\end{document}